# Pressure in active matter

Guo Yu[1], Ruiyao Li[1], Fukang Li[1], Jiayu Zhang[1], Xiyue Li[1], Zequ Chen[1], Joscha Mecke[1,2,*] and Yongxiang Gao[1*]

[1] Institute for Advanced Study, Shenzhen University, Shenzhen 518060, China
[2] College of Physics and Optoelectronic Engineering, Shenzhen University, Shenzhen 518060, China

E-mail: joscha.mecke@szu.edu.cn; yongxiang.gao@szu.edu.cn

## Abstract

In the last decade, the study of pressure in active matter has attracted growing attention due to its fundamental relevance to nonequilibrium statistical physics. Active matter systems are composed of particles that consume energy to sustain persistent motion, which are inherently far from equilibrium. These particles can exhibit complex behaviors, including motility-induced phase separation, density-dependent clustering, and anomalous stress distributions, motivating the introduction of active swim stress and swim pressure. Unlike in passive fluids, pressure in active systems emerges from momentum flux originated from swim force rather than equilibrium conservative interactions, offering a distinct perspective for understanding their mechanical response. Simple models of active Brownian particles (ABPs) have been employed in theoretical and simulation studies across both dilute and dense regimes, revealing that pressure is a state function and exhibits a nontrivial dependence on density. Together with nonequilibrium statistical concepts such as effective temperature and effective adhesion, pressure offers important insight for understanding behaviors in active matter such as sedimentation equilibrium and motility induced phase separation. Extensions of ABPs models beyond their simplest form have underscored the fragility of pressure-based equation of state, which can break down under factors such as density-dependent velocity, torque, complex boundary geometries and interactions. Building on these developments, this review provides a comprehensive survey of theoretical and experimental advances, with particular emphasis on the microscopic origins of active pressure and the mechanisms underlying the breakdown of the equation of state.

## Introduction

Over the past two decades, active matter has gained widespread prominence across disciplines such as colloid science[1, 2], condensed matter physics[3, 4], nonequilibrium thermodynamics[5, 6], and self-assembly[7-9]. When a system's components consume energy from the environment to generate forces and remain in a nonequilibrium state, it is referred as an active matter system[10-12]. Active matter offers a paradigm for numerous driven systems within the realm of soft and living matter, including biological systems such as bacterial colonies[13], tissues and cells[14], flocks of birds[15], schools of fish[16-18], bee colonies[19, 20], crowds of human being[21], and synthesized active matter consisting of various types of self-propelling particles, ranging in size from nanometers to millimeters[22, 23].

In active matter system, the combination of activity and interactions between individual constituents leads to the emergence of active stresses, which act in addition to the usual equilibrium stresses such as nematic or viscous stresses[4, 17, 24]. For example, in a bacterial colony of *E. coli*, the dynamics are driven by viscous, elastic, and active stresses, with bacterial hydrodynamic propulsion creating extensile flow fields that pull in fluid from the sides and push it out along the bacterial axis[25, 26]. In the absence of activity, the dynamics reduce to nematohydrodynamic behavior of nematic liquid crystals[27].

Following this pathway, it is tempting to define the trace of the total nonequilibrium stress tensor as an effective active pressure.

The physical meaning of such an effective active pressure intuitively can be understood as the pressure on the bondary to confine active particles in space[28, 29], analogous to the kinetic theory of gases, where collisions between molecules and container walls generate pressure, or similar to the Brownian osmotic pressure generated by molecules or colloidal solutes in a solution. However, a key feature of active matter is that it is inherently in a thermodynamic nonequilibrium state, making its collective behavior impossible to understand using conventional statistical mechanics[30, 31]. In equilbrium systems, pressure is a state function, and attains the same value whether derived from microscopic momentum flux, statistical thermodynamics or hydrodynamics. While some of these definitions are extended to define pressure in active matter systems, their convergence and the existence of an equation of state are by no means guaranteed[28, 32-34], which renders the understanding and physical interpretation of pressure in active matter a challenging problem.

The conceptual understanding of active stresses and effective pressure is still in early stages, with different perspectives developed. Considerable discussion has centered on the conditions under which active pressure can be regarded as a state function—governed by internal momentum flux[35] — and on the circumstances in which this description breaks down, causing the mechanical pressure on the confining wall to deviate from that derived from the bulk[32, 36-38]. Despite these challenges, active pressure remains to be a valuable concept for understanding collective behaviors such as self-assembly, phase separation, or pattern formation in active matter systems[13, 14, 17, 39-41], controlling self-propelled particles[37], as well as for designing, fabricating, and micromachines[42, 43].

In this paper, we conduct a synthesized review on past studies of active pressure, exploring the microscopic mechanical origins especially in active Brownian particles. We start from the simple ABPs model, demonstrating that in this system the active pressure behaves as a state function and exhibits a nontrivial dependence on density. As part of the broader effort to extend equilibrium concepts to nonequilibrium situations and thereby apply well-known results to complex systems, researchers have sought to explain the active pressure of simple ABPs by drawing analogies to pressure in equilibrium systems, introducing the concepts of effective temperature and effective adhesion. Going beyond the simplest ABP models has revealed that pressure-based equations of state are highly sensitive and can fail when influenced by factors such as density-dependent propulsion speeds, torques, and complex boundary shapes or interactions. Here we trace these developments, highlight the key mechanisms behind the breakdown of the equation of state, and discuss how these insights inform both theoretical modeling and experimental design in active matter research.

# 1. Pressure in equilibrium systems

## 1.1 Pressure in an ideal gas

In the ideal gas, the simplest and most widely used models in statistical mechanics, the pressure can be shown to arise from the collective impact of gas particles colliding with the walls of the container (Figure 1a). These collisions transfer momentum to the walls, and the frequency and strength of these collisions are directly related to the thermal energy of the particles, $k_B T$, where $k_B$ is the Boltzmann constant and $T$ is the absolute temperature. This leads to the well-known expression for the ideal gas pressure[44]:

$$P_{\text{id}} = nk_B T, \quad (1)$$

where $n = N/V$ is the number density, with $N$ the total number of gas molecules and $V$ the volume of the container.

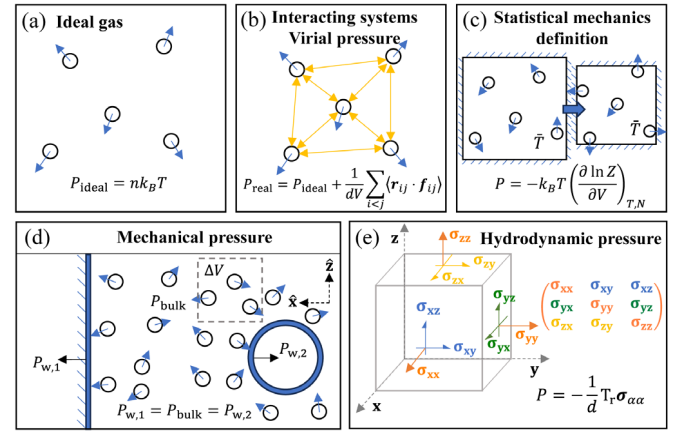

**Figure 1. Different definitions of pressure in equilibrium systems.** (a) Pressure in an ideal gas. (b) Virial pressure contributed by intermolecular interactions. (c) Pressure defined based on statistical mechanics. (d) Mechanical pressure defined as the force per unit area on the confining boundaries or walls; $P_{\text{w},1}$ and $P_{\text{w},2}$ represent the mechanical pressure on a flat wall and a curved wall, respectively. The pressure is uniform throughout the system, such that $P_{\text{w},1} = P_{\text{w},2} = P_{\text{bulk}}$. (e) Pressure is also defined as the sum of the trace of the hydrodynamic stress tensor.

## 1.2 Virial pressure due to interactions between particles

For realistic systems, interactions between particles also contribute to the pressure in the system (Figure 1b), which can be calculated via the Virial theorem[45]. It provides a foundational route to define pressure in many-body systems from a microscopic perspective, relating the average force acting between particles to the macroscopic pressure via the spatial distribution of particles and their mutual interactions. For a system of $N$ particles confined in a volume $V$ in $d$ dimensions, the pressure can be written as the following,

$$P_V = \frac{1}{dV}\sum_{i<j}\langle \vec{r}_{ij}\cdot\vec{f}_{ij}\rangle, \quad (2)$$

where $\vec{f}_{ij}$ and $\vec{r}_{ij}$ denote the force and relative distance between two particles $i$ and $j$, respectively. Here, the dot product quantifies how pairwise forces contribute to the net momentum exchange and hence to the pressure. This formulation offers an intuitive way to compute pressure in molecular simulations. The concept is not only applicable to gases, but also can be applied to suspensions of Brownian particles where entropy drives the colloids to spread out leading to an osmotic pressure analogous to the reasoning above[46].

*1.3 Statistical definition*

In statistical mechanics, macroscopic thermodynamic quantities can be derived from the microscopic properties of the system through the partition function $Z$, which encodes the statistical weight of all possible microstates[47]. Take canonical ensemble as an example, the partition function is defined as,

$$Z = \int e^{-H(p,q)/k_B T}\,\mathrm{d}p\,\mathrm{d}q, \quad (3)$$

where $H(p,q)$ is the Hamiltonian of the system, p and q represent momentum and coordinate, respectively. In this case, pressure can be expressed as a derivative of partition function with respect to volume,

$$P = -k_B T \left(\frac{\partial \ln Z}{\partial V}\right)_{T,N}. \quad (4)$$

This expression conveys the fundamental thermodynamic definition of pressure, relating it to changes in the system's accessible microstates and entropy with volume (Figure 1c), i.e. pressure is understood to originate from the volume dependence of the partition function, rather than from a purely mechanical or kinetic description.

*1.4 Mechanical definition*

Generally, pressure is understood as the normal force exerted per unit area on a surface[48]. This mechanical definition holds regardless of whether the system is in equilibrium or not (Figure 1d):

$$P = \frac{F}{A}, \quad (5)$$

where $F$ is the total normal force and $A$ is the total surface area over which the force is applied.

By making the dependence of the thermodynamic free energy on the boundary position explicit and taking its derivative, the resulting term can be identified as the force exerted on the particles by the wall, thus connecting the thermodynamic expression of pressure to its mechanical wall-force form[32]:

$$P = \int_0^\infty n(x)\partial_x V(x)dx, \quad (6)$$

where $V(x)$ is a wall potential. This equation still holds even if the system contains other types of particles (such as solvent molecules), as long as those particles do not exert any direct force on the wall (*i.e.*, the wall is semipermeable to them)[32]. In this case, the pressure $P$ is identified as the osmotic pressure.

*1.5 Hydrodynamic definition*

In hydrodynamics, pressure arises as a macroscopic, continuum field that captures the isotropic part of the stress tensor in a fluid[48]. The pressure is defined as the sum of the trace of stress tensor $\boldsymbol{\sigma}$ divided by the dimension $d$[49],

$$P = -\frac{1}{d}\sum_{\alpha=1}^{d}\sigma_{\alpha\alpha}. \quad (7)$$

Note that it is fundamentally different from the microscopic Virial expression, but connected through coarse-graining.

## 2. Pressure in active systems

As active matter consumes energy from the environment, they are inherently driven out of equilibrium[17]. The question then is how to define pressure in active systems. To make progress, ABPs models have been used in simulations and theoretical calculations due to their simplicity. It is generally accepted that pressure in active systems contains two contributions[28, 32-34, 50-53],

$$P = P_p + P_{\text{swim}}. \quad (8)$$

One is the common pressure experienced in a passive system, denoted as $P_p$, and the other is contributed by the self-propelled motion of particles, known as swim pressure[28], $P_{\text{swim}}$. The passive part of the pressure can be further expressed as the sum of two terms,

$$P_p = P_{\text{id}} + P_V, \quad (9)$$

with $P_{\text{id}}$ and $P_V$ denotes ideal-gas like contribution from random Brownian motion of particles and the Virial contribution arising from mutual interactions.

In the past decade, various approaches to derive expressions for pressure in active matter have been explored[28, 32, 34, 40, 50, 51, 54, 55]. Though pressure is a state function in equilibrium systems, it is far from clear what to expect in active matter due to their far-from-equilibrium nature. It becomes even more complicated in the presence of various confining boundaries[50], which may alter the spatial distribution and dynamics of particles. Indeed, it has prompted extensive discussion on whether, and under what conditions, pressure qualifies as a state function, thereby providing an opportunity to further develop the statistical physics of nonequilibrium systems.

*2.1 Active Brownian Particles*

Active Brownian particles are a class of models composed of particles that possess self-propulsion[56]. The propulsion speed and direction are influenced by noise, friction, and external fields, where the friction coefficient $\gamma(\mathbf{r},t)$ may depend on the particle's position and velocity, and can even take negative values to model the injection of energy from an external pump to sustain self-propelled motion of particles. The particle's orientation evolves over time, which can be driven by rotational Brownian motion, deterministic torques, or a combination of both. ABPs model "dry" active systems without hydrodynamic coupling. Our discussion mainly revolves around this model to explore the conditions under which an equation of state holds and the situations where it breaks down (Figure 2).

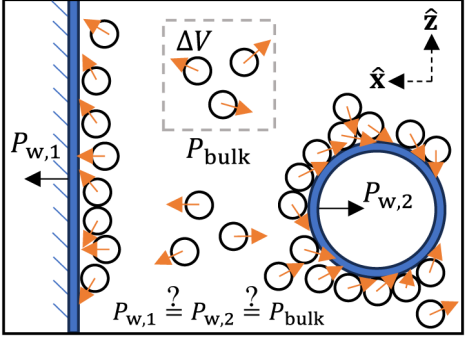

**Figure 2. Pressure in ABP model systems.** $P_{w,1}$ and $P_{w,2}$ are mechanical pressure at flat or curved walls, respectively. $P_{bulk}$ represents bulk pressure away from boundaries in active system.

*2.2 Irving–Kirkwood formalism for stress tensor*

The microscopic definition of the stress tensor in statistical mechanics is commonly formulated using the Irving–Kirkwood (IK) formalism, first developed in 1950.[57] In this framework, the stress tensor is defined as the flux of linear momentum across a surface element, derived from the conservation laws applied at the microscopic scale. It contains contributions from both kinetics and configurations, representing the momentum transport due to particle motion and interparticle interactions, respectively. The IK stress tensor for a system of $N$ particles interacting via pairwise forces can be expressed as:

$$\boldsymbol{\sigma}(\mathbf{r},t) = -\sum_i m_i \mathbf{v}_i \mathbf{v}_i \delta(\mathbf{r}-\mathbf{r}_i)$$
$$-\frac{1}{2}\sum_{i\neq j} \mathbf{r}_{ij} \mathbf{F}_{ij} \int_0^1 \delta(\mathbf{r}-\mathbf{r}_i + \lambda \mathbf{r}_{ij})\,d\lambda \quad (10)$$

Here, $\mathbf{r}_{ij} = \mathbf{r}_i - \mathbf{r}_j$, $m_i$ is the mass of particle $i$, and $\mathbf{F}_{ij}$ is the force exerted on particle $i$ by particle $j$. The first term represents the momentum carried by the particles. The second term accounts for the contribution from interparticle forces, integrated along the line connecting particles $i$ and $j$, with the force contribution distributed uniformly over the entire segment.

*2.3 Swim pressure as the Trace of swim stress tensor*

The IK formalism provides a well-established framework for defining the stress tensor in particle-based systems. To describe stress in active matter, the IK formalism has been further extended to incorporate nonequilibrium contributions from active forces.

To put into account the contribution from persistent propulsion of active particles, swim stress for overdamped systems is defined in the spirit of the IK formalism[28, 34],

$$\boldsymbol{\sigma}_{\text{swim}} = n\langle \mathbf{r}\mathbf{F}_{\text{swim}}\rangle. \quad (11)$$

In this case, the force is replaced by the effective self-propulsion force $\mathbf{F}_{\text{swim}}$.

While the swim stress provides a tensorial description of momentum flux, the pressure can be obtained by taking its trace[57]. That is, swim pressure corresponds to the sum of the diagonal components of the swim stress tensor,

$$P_{\text{swim}} = -\frac{1}{d}\sum_{\alpha=1}^{d} \sigma_{\alpha\alpha}^{swim}. \quad (12)$$

This scalar definition of swim pressure as the trace of the swim stress tensor can also be understood from a complementary perspective: an extension of the Virial theorem[45]. Swim pressure has been recast in a Virial-like form involving the dot product of the self-propulsion force and particle position[51, 58, 59],

$$P_{\text{swim}} = -\frac{1}{dV}\sum_i^N \langle \mathbf{F}_i^{\text{swim}} \cdot \mathbf{r}_i \rangle. \quad (13)$$

Here, conservative forces beween particles are replaced by swim forces, in which case the Virial theorem remains a useful framework for interpreting active pressures in nonequilibrium systems[28].

### 3. Pressure is a state function in specific ABPs models

*3.1 Overdamped simple ABPs model*

To understand the emergence of active stresses and pressure in nonequilibrium systems, researchers have focused on simple overdamped active Brownian particles (simple ABPs) systems. In this model, active particles evolve via a Langevin equation of motion,

$$\zeta \dot{\mathbf{r}}_i(t) = \zeta U_0 \mathbf{n}_i(t) + \mathbf{F}_i + \boldsymbol{\xi}_i(t). \quad (14)$$

Where $\mathbf{r}_i(t)$ is the particle location, $U_0$ is the constant swimming speed, $\mathbf{n}_i(t)$ is the unit vector along the axis of self-propulsion, $\zeta$ is the friction from the suspending fluid, $\boldsymbol{\xi}_i(t)$ is the thermal random force and $\boldsymbol{F}_i$ is the total force on the particle. The random force satisfies $\langle \boldsymbol{\xi}_i(t) \rangle = 0$ and $\langle \boldsymbol{\xi}_i(t) \cdot \boldsymbol{\xi}_i(0) \rangle = 2dk_BT\delta(t)$. The active swim force takes the following form,

$$\boldsymbol{F}_{\text{swim}} = \zeta U_0 \boldsymbol{n}(t). \quad (15)$$

Consider the position evolution equation (taking the case without thermal noise as an example):

$$\boldsymbol{r}(t) = \int_0^t U_0 \boldsymbol{n}(t')dt'. \quad (16)$$

Substituting Eq. (15) and Eq. (16) into Eq. (13), using the rotational autocorrelation function of active Brownian particles, we then can get the swim pressure in the function of which we will discuss in section 3.1.1.

*3.1.1 Pressure in overdamped simple ABPs models*

For an active Brownian particle moving at low Reynolds number, the time autocorrelation of its orientation unit vector $\boldsymbol{n}(t)$ is

$$\langle \boldsymbol{n}(t) \cdot \boldsymbol{n}(0) \rangle = e^{-D_R t}, \quad (17)$$

where $D_R$ is the rotational diffusion coefficient. This exponential decay leads to the following relation:

$$\langle \boldsymbol{n}_i \cdot \dot{\boldsymbol{r}}_i \rangle = -\langle \dot{\boldsymbol{n}}_i \cdot \boldsymbol{r}_i \rangle. \quad (18)$$

On the other hand, the mean-square rate of change of $\boldsymbol{n}_i$ is governed by rotational diffusion $\langle \dot{\boldsymbol{n}}_i \cdot \boldsymbol{r}_i \rangle = -D_R \langle \boldsymbol{n}_i \cdot \boldsymbol{r}_i \rangle$, which give $\langle \boldsymbol{n}_i \cdot \dot{\boldsymbol{r}}_i \rangle = D_R \langle \boldsymbol{n}_i \cdot \boldsymbol{r}_i \rangle$. Combining above with Eq (13), $P_{\text{swim}}$ can be expressed as the following,

$$P_{\text{swim}} = -\frac{n}{dD_R}\zeta U_0 U_\phi. \quad (19)$$

Where $U_\phi$ represents the average particle velocity projected along its swimming direction at various packing fraction $\phi$,

$$U_\phi = \frac{1}{N}\sum_i^N \langle \boldsymbol{n}_i \cdot \dot{\boldsymbol{r}}_i \rangle. \quad (20)$$

Here we arrived at a general formula for swim pressure in overdamped simple ABPs. One velocity factor $U_0$ stays the same, and the other is replaced by a density-dependent speed $U_\phi$. Here $\phi = nv_p$, the volume fraction in 3d, and $\phi = n\sigma_p$, the area fraction in 2d, where $v_p$ and $\sigma_p$ are the volume and area of a particle, respectively. Physically, $U_\phi$ measures how much of the self-propulsion is preserved in the actual motion of the particles, after accounting for collisions and interactions. It serves as a key kinetic parameter in the virial expression for swim pressure[50, 51, 55, 58].

In the dilute regime, $U_\phi = U_0$. The swim pressure given in Eq. (19) then reduces to:

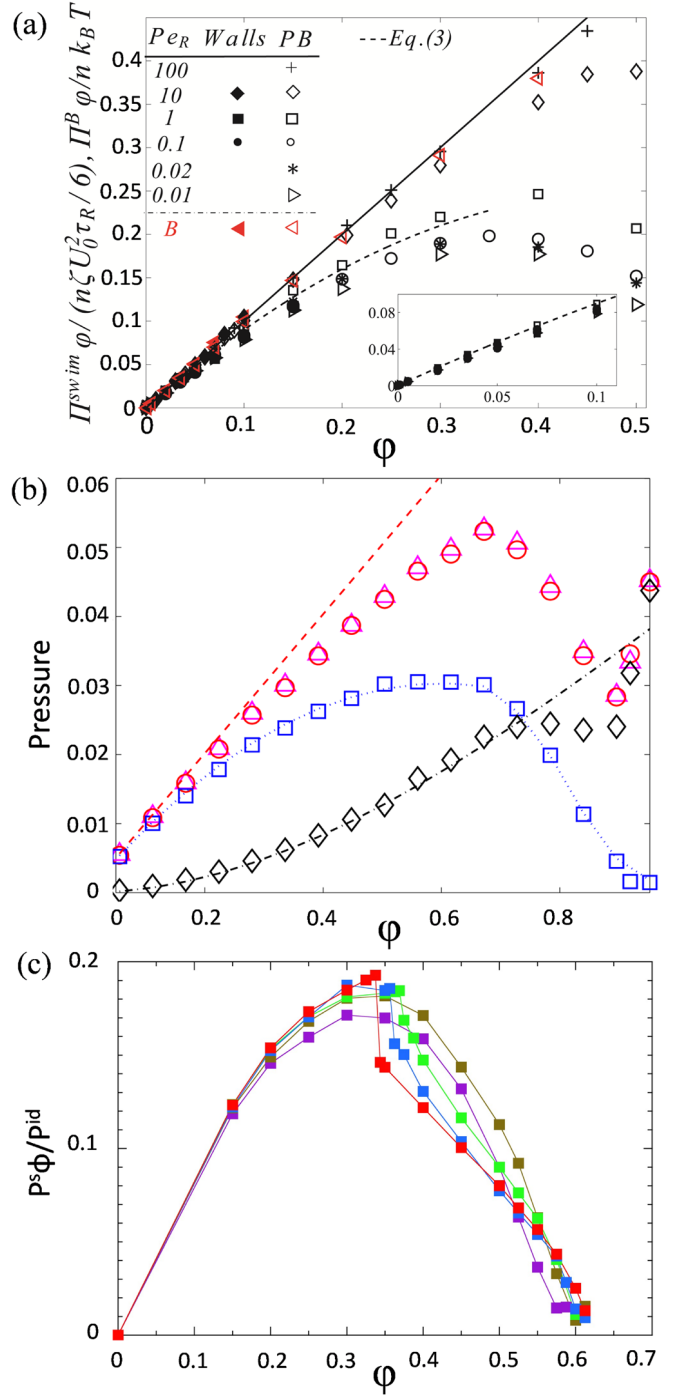

**Figure 3. The dependence of simple ABPs' pressure on packing fraction.** (a) The swim pressure at different reorientation $Pe_R$ with periodic boundary conditions and walls. Reproduced from Ref.[28], © APS, used with permission. (b) The total pressure calculated from the IK formula (triangles) and the mechanical pressure on the walls (circles). Reproduced from Ref.[34], with permission from The Royal Society of Chemistry. (c) Swim pressure of ABPs at various Péclet numbers (Pe = 9.8 (purple), 29.5 (olive), 44.3 (green), 59.0 (blue), and 295.0 (red)). Reproduced from Ref.[51]. © The Royal Society of Chemistry, CC BY 3.0.

$$P_{\text{swim}}^0 = \frac{\zeta n U_0^2}{d D_R}, \quad (21)$$

where this result has been reported by multiple studies[28, 34, 51] for both two and three dimensional systems. In both cases, the pressure introduced by activity is proportional to $U_0^2$, revealing the fundamental origin of active pressure (Figure 3).

In dense simple ABPs systems, the swim pressure is decreased because the orientation of the active particles changes on the timescale $\tau_R$ such that the particles do not necessarily reorient when colliding and thus mutually obstructing the active dynamics[28, 34, 50, 51, 55, 58, 60]. That is, $U_\phi$ no longer equal to $U_0$ but instead decreases[34],

$$U_\phi = U_0(1 - \lambda \phi). \quad (22)$$

Where $\lambda$ is a constant parameter, characterizing the slope of the decrease in particle propulsion speed with packing fraction $\phi$. As shown in Figure 3b, the dashed magenta line represents the calculated ideal-gas pressure without any fitting parameters, while the magenta symbols deviate from it.

By using concepts from active microrheology, the leading-order correction predicts a linear decrease of the swim pressure with increasing density[28] (Figure 3a):

$$\frac{P_{\text{swim}}}{P_{swim}^0} = 1 - \phi. \quad (23)$$

From the interparticle forces (steric repulsion), containing leading order linear increase of the pressure with the colloidal density. The full active pressure to leading order in density is thus obtained as:

$$\frac{P_{\text{act}}}{P_{swim}^0} = 1 - \phi(1 - 3\text{Pe}_R) + \mathcal{O}(\phi^2). \quad (24)$$

This equation can be understood as an expansion of an equation of state of active matter in terms of deviations about the effective ideal gas state. The negative term (second term of right-hand side of the equation) plays a similar role as the second-virial coefficient of equilibrium colloidal systems, and when it is negative, it indicates effective attractive interactions between the active particles[61].

In Gompper's study[51], the pressure collapses onto a common curve described by:

$$\frac{P}{P_{swim}^0} = \phi(1 - \kappa \phi), \quad (25)$$

with $\kappa$ of order unity. This form captures the ideal-gas-like linear growth at low $\phi$ and the quadratic suppression at higher densities, reflecting the onset of interaction effects in active systems. On the other hand, the reorientation Péclet number $\text{Pe}_R = a/(U_0 \tau_R)$ controls the reorientation frequency increases together with the reorientation frequency. Accordingly, for high $\text{Pe}_R$, the effective ideal-gas active swim pressure is obtained in simulations even at high densities[28] (Figure 3 c).

### 3.1.2 Effective temperature in the dilute regime

The relationship between swim pressure and density in dilute situation has been explored in active colloidal suspensions through sedimentation experiments[54]. To further characterize the thermodynamic properties of such self-propelled systems, a method based on the equilibrium fluctuation–dissipation theorem has been proposed for defining an effective temperature[62]. Félix Ginot and his colleagues[54] designed the experimental setup shown in the inset of Figure 4a. They tilted the system by a small angle $\theta$ along the $z$ direction to form a continuous two-dimensional monolayer of gold-platinum Janus microspheres immersed in a hydrogen peroxide bath at the bottom of the observation chamber, establishing varying packing fractions along the $z$ axis by balancing the osmotic pressure with the gravitational potential. The simulation results also agree well with the experimental results (Figure 4b).

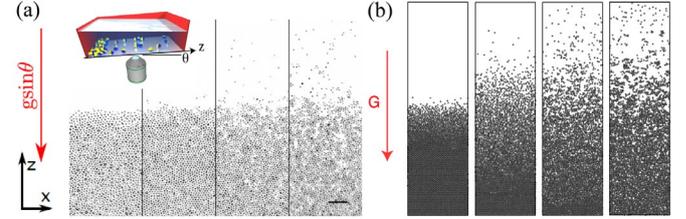

**Figure 4. Sedimentation equilibrium of self-propelled particles in experiments and simulations.** (a) Sedimentation profiles of phoretic gold-platinum Janus colloids; Activities increase from left to right. Reproduced from Ref.[54], © Félix Ginot, published by APS, [CC BY 3.0.](b) Simulation snapshots for self-propelled hard disks under gravity; activities increase from left to right. Reproduced from Ref.[54], © Félix Ginot, published by APS, [CC BY 3.0.]

In the long-time limit of the ABPs dynamics, the particles can be regarded as diffusing with an activity-induced effective diffusion coefficient[63, 64] $D_{\text{eff}} = U_0^2 \tau_R$. In dilute situations, the steady-state distribution in the gravitational potential follows a Boltzmann distribution thermalized at $T_{\text{eff}}$:

$$\mathcal{P}(z) \propto \exp\left[-\frac{mgz \sin\theta}{k_B T_{\text{eff}}}\right], \quad (26)$$

which defines an effective temperature.

Moreover, it has also been shown experimentally that in dilute suspensions an effective Stokes-Einstein relation can be established[54] stating that $k_B T_{\text{eff}}/\zeta = D_{\text{eff}}$, thus implying $k_B T_{\text{eff}} = U_0^2 \tau_R \zeta$. The swim pressure of dilute simple ABPs thus can be described by effective temperature:

$$P_{\text{swim}} = n k_B T_{\text{eff}}. \quad (27)$$

On the use of the Einstein relation, Solon *et al.*[32] also arrived at the same conclusion by setting the model as torque-free (for example, spherical) particles:

$$P = \frac{\zeta n_0 U_0^2}{2(D_R + \alpha)} + n_0 k_B T = n_0 k_B T_{\text{eff}}. \quad (28)$$

The effective temperature $T_{\text{eff}}$ is related to the actual temperature $T$ by[54, 65] as shown in the Figure 5:

$$\frac{T_{\text{eff}}}{T} = \left(1 + \frac{2}{9}\text{Pe}^2\right). \quad (29)$$

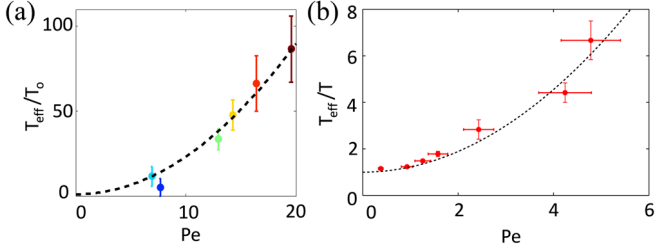

**Figure 5.** (a) Raw effective temperature $T_{\text{eff}}$ as a function of the Péclet number Pe, exhibiting a quadratic scaling trend. The data are obtained for self-propelled Janus colloids and directly compared with theoretical predictions. Reproduced from Ref.[54], © Félix Ginot, published by APS, CC BY 3.0. (b) Normalized effective temperature $T_{\text{eff}}$ versus Pe for sedimenting light-activated colloids, demonstrating quantitative agreement with the theoretical model across a broad range of activity strengths. Reproduced from Ref.[65], Copyright (2010) by the American Physical Society.

Overall, experimental and simulation studies collectively indicate that, in the dilute limit, systems composed of simple ABPs exhibit behaviors analogous to those of an ideal gas. In particular, these studies show that such active systems can be effectively described using the concept of an activity-dependent effective temperature $T_{\text{eff}}$, which captures how self-propulsion modifies their thermodynamic and mechanical properties compared to equilibrium systems. This analogy allows researchers to employ familiar equilibrium statistical mechanical tools, offering valuable insights into understanding the fundamental physics governing dilute active matter systems.

It is worth noting that, if the Brownian motion of the particles stemming from the thermal agitation of the active colloids is much stronger than the active contributions, the contributions of the effective active pressure will be hidden in the thermal equilibrium dynamics of the colloids[28]. The importance of the thermal contributions is typically measured with the translational Peclét number $\text{Pe}_s = U_0 a/D_0$. Measurements are more precise[54] for $\text{Pe}_s$ above 10, since the thermal agitation cannot be turned off in experiments, this is a crucial lever for experimental design.

### 3.1.3 Effective adhesion at finite densities

The existence of swim pressure has been explicitly confirmed in a suspension of self-propelled Janus colloids subject to an acoustic confinement of size much larger than the colloids themselves[66]. These self-propelled particles swim via self-diffusiophoresis[67] in a hydrogen peroxide solution. Adding a harmonic force $k(\boldsymbol{r}(t) - \boldsymbol{r_0})$ leads to a balance between the active and the confinement forces, where $k$ is the spring constant of acoustic confinement, $\boldsymbol{r}(t)$ is the particle position at time $t$. Upon sudden release, the system exhibited a "crystal explosion" (Figure 6a), reflecting the tendency of active particles to escape in the absence of confinement. This observation directly illustrates the physical origin of the swim pressure, which the mechanical pressure exerted by active particles as a result of their self-propulsion[28, 29]. Furthermore, the pressure decreases with increasing trap stiffness $k$ (Figure 6b). The strong trap reduces the "moment arm", defined as the path length that active particles travel before reorienting. Such confinement thus results in a decreased swim pressure. At finite densities, particle obstructions and collisional slow-down effects similarly shorten the moment arm, causing the swim pressure to decrease further as the density increases.

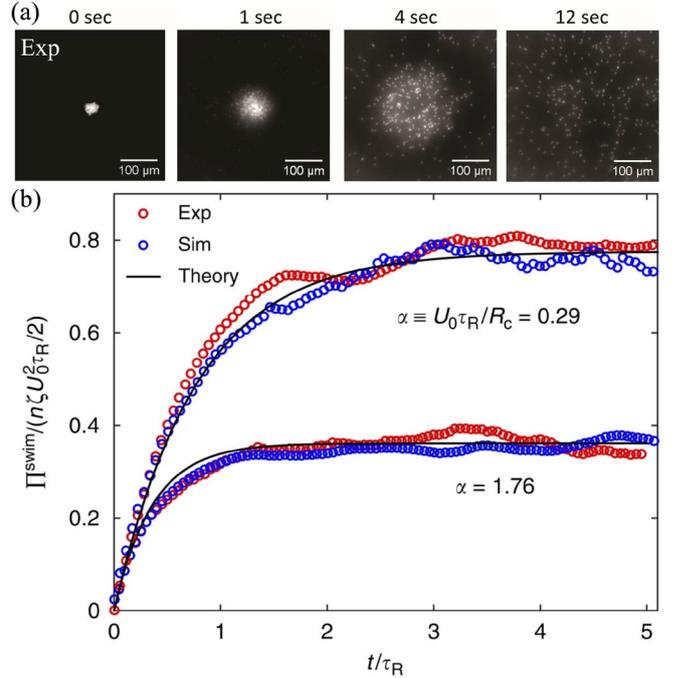

**Figure 6. Crystal explosion experiment.** (a) Experimental observation of an active crystal rapidly expanding after acoustic trap released, showing a transition from a dense cluster to a diffusive steady state. Reproduced from Ref.[66] under CC BY 4.0. (b) Swim pressure of Janus particles under different confinement strengths ($\alpha = U_0 \tau_R/R_C$), where $R_C = \zeta U_0/k$ is the trap size. Red and blue dots show experimental and simulation results, and black lines are theoretical predictions. Reproduced from Ref.[66] under CC BY 4.0.

At high particle density, an effective attractive interaction or adhesion emerges between active particles, from which an effective adhesion strength could be extracted. Based on this, a unique scaling law relating activity to self-propulsion was identified in both sedimentational experiments and simulations[54]. At finite densities, which had not been systematically explored in experiments before, both experimental and numerical results show that the functional form of the equation of state changes continuously with increasing activity. This change cannot be fully explained by introducing an effective temperature as in section 3.1.2, highlighting the limitations of this approach in describing active systems.

The behavior of the active system was effectively mapped onto the Baxter model[68], which describes an equilibrium fluid with adhesive interactions. In this model, particles interactions are described by a square-well potential consisting of a hard-core repulsion at short distances and a narrow attractive well just beyond the particle diameter. The attractive interaction is short-ranged, and its strength is quantified by a dimensionless adhesion parameter $A$. The pressure-density relation can then be expanded as

$$Z = \frac{P}{nk_BT} = 1 + b_1\phi + b_2\phi^2 + \mathcal{O}(\phi^3). \quad (30)$$

The first two virial coefficients are known analytically[69], with $b_1 = 2 - A$, $b_2 = \frac{25}{8} - \frac{25}{8}A + \frac{4}{3}A^2 - 0.122A^3$. That is, when $A$ increases, the pressure in units of the ideal gas pressure changes from the monotonic hard-particle behavior to a nonmonotonic density dependence for $A > 2$ as the initial slope given by $b_1$ then becomes negative. The effective adhesion $A$ increases with self-propulsion, as a function of effective temperature[54] (Figure 7):

$$A \sim \sqrt{T_{\text{eff}}/T}. \quad (31)$$

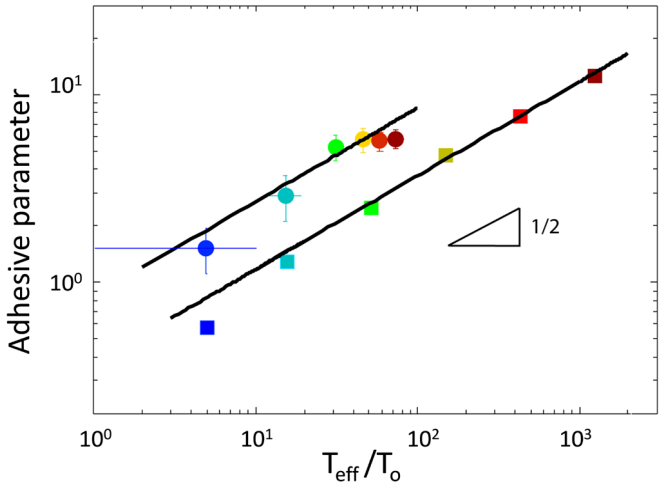

**Figure 7.** Adhesive parameter $A$ as a function of $T_{\text{eff}}/T_0$, where $T_0$ is real temperature. showing scaling $A \sim \sqrt{T_{\text{eff}}/T_0}$ in both experiments (circles) and simulations (squares). Solid lines indicate the expected slope of 1/2 with different prefactors. Reproduced from Ref.[54]. © 2015 American Physical Society, CC BY 3.0.

The pressure in active systems remains describable by an equation of state even after introducing the effective adhesion parameter $A$. This indicates that the incorporation of effective adhesive interactions among active particles does not compromise the thermodynamic consistency of the system. Instead, it expands the applicability of the equation of state, enabling it to account for more complex phenomena such as clustering and motility-induced phase separation.

*3.1.4 Motility induced phase separation*

In ABPs systems, motility induced phase separation (MIPS) arises from self-propelled particles accumulation in regions where they move more slowly[70], which may happen through direct pairwise forces collision[71-73] or density-dependent propulsion speed like quorum sensing[74-78]. When the stochastic density is coarse-grained over time and space, one obtains an equation for the mean density field, *i.e.*, the hydrodynamic equation, which corresponds to the deterministic limit. In the uniform limit, the generalized pressure exactly recovers the EOS reported in Ref. [50].

The coexistence of dilute and dense phases in an ABP system can occur even in the absence of attractive interactions[5, 71, 72, 79]. In a particular range of Peclét numbers and the active particle densities, this can lead to the formation of a dense phase surrounded by an active gas phase[71, 72, 80]. For low $Pe_R$ and finite $\phi$, a bimodal distribution of the packing fraction is obtained, indicating phase coexistence of an active gas and a cluster phase[5, 73, 81].

Such phase coexistence behavior can be explained to large extent by passive particles with attractive interactions. The Baxter model equilibrium virial equation of state for adhesive particles can be used in order to recover the adhesive behaviors in ABPs systems observed in simulations and experiments[54]. Moreover, the non-monotonic behavior in Eq. (24) has been reproduced in simulations, where the total pressure at $\phi = 0.6$ is lower than at $\phi = 0.3$ for a value of $Pe_R \approx 0.03$, *i.e.*, when the second term on the right-hand side of Eq. (24) is negative[28].

Systematic studies[5, 58, 80, 82] on the pressure behavior of spherical ABPs undergoing MIPS, with a particular focus on whether an equation of state can describe such nonequilibrium transitions reveals that the liquid-gas phase coexistence and phase transition in ABPs systems is consistent with equilibrium first-order phase transitions, characterized by a flat pressure plateau between coexisting phases. It suggests that pressure equalization occurs once nucleation has taken place. When nucleation is suppressed (*e.g.*, in open systems without walls), the system remains in a metastable homogeneous state[58]. No Maxwell construction can be applied and large structural differences between the coexisting phases lead to a high nucleation barrier[58, 80]. However, the effective active pressure can be determined by bulk correlators, showing that the pressure is not determined by the interactions with the walls, similar as in equilibrium systems. Additionally, the interfacial tension between the two phases is very large (compared to equilibrium systems) and negative, which would imply an energy reward if the interface is increased[81, 82]. However, this work is not released to the system, but it is spent by the active particles to drive the surrounding fluid, thus stabilizing the cluster phase[82].

These findings underscore the fundamental differences between MIPS in active systems and phase separation in equilibrium fluids, especially in the role of interfacial contributions and the breakdown of classical thermodynamic constructs such as the Maxwell construction. They also highlight the need for further theoretical and experimental work to understand how boundary conditions, interfacial properties, and nucleation dynamics interact to determine the phase behavior of active matter. Such a mechanical perspective forms the foundation for investigating whether an equation of state exists in active systems. This interpretation remains valid even under nonequilibrium conditions, provided the pressure is derived directly from microscopic force contributions[28].

*3.2 Underdamped simple ABPs model*

To extend the applicability of active pressure theories to mesoscopic and macroscopic regimes, it is necessary to account for inertial effects in active particle models. The underdamped simple ABPs model introduces finite mass $m$ into the description of torque-free, spherical, self-propelled particles, thereby retaining both translational and rotational inertia in the particle dynamics.

The translational motion is governed by the Langevin equation with inertia:

$$m\ddot{\boldsymbol{r}}_i(t) = -\zeta \dot{\boldsymbol{r}}_i(t) + \zeta U_0 \boldsymbol{n}_i(t) + \boldsymbol{F}_i + \boldsymbol{\xi}_i(t). \quad (32)$$

The orientation dynamics remains overdamped and follows the same rotational diffusion as in the overdamped case.

This underdamped case allows for the analysis of stress generation under finite inertia, particularly relevant when the Stokes number $St_R = (m/\zeta)/\tau_R$ becomes non-negligible. Compared to the overdamped case, the underdamped model captures key physical phenomena such as delayed reorientation, finite response times, and the emergence of Reynolds stress, all of which significantly affect the definition and measurement of active pressure, especially in confined or dense systems.

*3.2.1 Swim–Reynolds pressure compensation in dilute regime*

To generalize the active pressure theory in order to understand more generic systems, it is necessary to incorporate inertial effects, particularly at mesoscopic or macroscopic scales. Based on a microscopic Irving–Kirkwood formulation, Ref.[83] extended the stress tensor expression of ABPs with finite mass, establishing the following decomposition:

$$\boldsymbol{\sigma} = \boldsymbol{\sigma}_{\text{kinetic}} + \boldsymbol{\sigma}_{\text{swim}} + \boldsymbol{\sigma}_{\text{V}}. \quad (33)$$

Here, the kinetic stress $\boldsymbol{\sigma}_{\text{kinetic}}$ (also called Reynolds stress) accounts for momentum transport due to inertial motion, while the swim stress reflects the nonequilibrium momentum flux originating from self-propulsion[33]. The interaction stress represents the classical virial contribution from interparticle forces.

Importantly, it was shown that inertia suppresses the swim stress due to the lag between the propulsion direction and actual particle velocity, which also be discussed in Ref.[84] that the inertial lag between particle orientation and direction of movement reduces the correlation $\langle \boldsymbol{r}\, \boldsymbol{F}_{\text{swim}} \rangle$ between the "momentum arm" $\boldsymbol{r}$ and the swim force $\zeta U_0 \boldsymbol{n}(t)$, which explicitly depends on the particle orientation. This effect is captured by the translational Stokes number $St_R = (m/\zeta)/\tau_R$, which measures the ratio between inertial relaxation time and rotational diffusion time. The swim stress diminishes as:

$$P_{\text{swim}} \propto \frac{1}{1 + 2St_R}. \quad (34)$$

But at the same time, inertia introduces an additional kinetic stress (Reynolds stress) term, consisting of the Brownian osmotic stress and a contribution stemming from self-propulsion which depends on $St_R$. Interestingly, the decrease in swim stress is exactly compensated by the increase in kinetic stress (Reynolds stress), so that the total pressure remains invariant with respect to inertia in dilute systems. This compensation indicates that the equation of state still holds for underdamped ABPs in the dilute regime, despite the redistribution of momentum flux among different stress contributions.

Furthermore, a conceptual revision of the swim stress formulation has revealed deeper insights into the distinction between local and global pressure definitions in underdamped models[35]. Early studies on overdamped ABPs concluded that the bulk pressure of ideal active particles is equivalent to that of passive gases, as the swim stress was thought to vanish in the bulk[55], which led to the belief that active pressure arises solely from particle-wall interactions. However, this conclusion does not hold in general condition, such as the presence of interia. The conventional formulation of swim stress vanishes locally and thus fails to capture internal momentum transfer. In contrast, the inertial-based formulation gives a finite local swim stress, which aligns with the previous discussion on Ref. [83, 84].

*3.2.2 Modified EOS at finite concentration*

At finite concentrations, the total pressure of active systems can still be described by a modified mechanical pressure theory, provided that both the Reynolds stress and interparticle interaction pressure are explicitly included[83, 84]. In the high-density regime, active particles exhibit pronounced phase separation and self-assembly behaviors, and their mechanical response becomes increasingly complex. Simulations reveal that both the swim pressure and Reynolds pressure decrease with increasing particle volume fraction $\phi$, primarily due to frequent collisions that limit self-propulsion and reduce the correlation in particle motion. Specifically, the swim pressure exhibits a concentration dependence of:

$$P_{\text{swim}} \propto \frac{1-\phi-\phi^2}{1+2\text{St}_R}, \quad (35)$$

while the Reynolds pressure follows a similar form,

$$P_{\text{Rey}} = nk_S T_S \frac{1-\phi-\phi^2}{1+1/2\text{St}_R} + nk_B T. \quad (36)$$

Where $k_S T_S$ denotes swim "energy scale", defined in three dimensions as $k_S T_S \equiv \zeta U_0^2 \tau_R / 6$. Notably, the sum of swim and Reynolds pressures, *i.e.*, the total active pressure, remains independent of the Stokes number $\text{St}_R$ at fixed Péclet number, but exhibits a strong dependence on concentration $\phi$. In addition, repulsive interactions between particles introduce a non-negligible interparticle (collisional) pressure, which increases monotonically with $\phi$ and contributes to the stabilization of the system.

Therefore, these results emphasize the importance of inertial effects in capturing the true contribution of active forces to the internal stress of the system, and further demonstrate that the pressure in underdamped simple ABPs also satisfies equation of state.

## 4. Pressure is not a state function in generic active systems

The validity of an equation of state in underdamped active matter systems is highly sensitive to system details. In dilute spherical active Brownian particle systems without torque, previous studies have shown that pressure remains a state function even when particle inertia is finite[83, 84]. While the equation of state may hold for simple, dilute, and torque-free ABPs, it does not apply to more complex active systems[32, 38, 85, 86].

### 4.1 Torque between particles

Interparticle alignment torque is one of the most extensively studied interactions in active matter and is a key factor in breaking the equation of state[17]. A key mechanism breaking the equation of state in active matter is torque between particles from non-central, anisotropic interactions[32].

The orientation of each ABP evolves under an aligning torque generated by neighboring particles, with strength set by the rotational mobility[32]. This torque depends explicitly on the relative positions and orientations of particles, making the interaction anisotropic and non-central. As a result, the particle distribution becomes sensitive to the wall potential, and the mechanical pressure depends not only on bulk variables but also on the microscopic details of the alignment rule and boundary geometry.

For example, in the visual perception model of Bechinger *et al.*[87], particles sense neighbors within a finite visual cone and adjust propulsion speed according to perceived density. This anisotropic sensing induces an effective alignment torque, which becomes non-reciprocal for asymmetric cones ($\alpha < \pi$). As a result, mechanical pressure depends on the specific interaction rule and boundary geometry, rather than solely on bulk variables. Simulations confirm that for asymmetric vision, steady-state pressure varies with cone aperture and cannot be inferred from bulk measurements, demonstrating that torque-mediated reorientation fundamentally disrupts the EOS in active matter.

### 4.2 Density-dependent propulsion

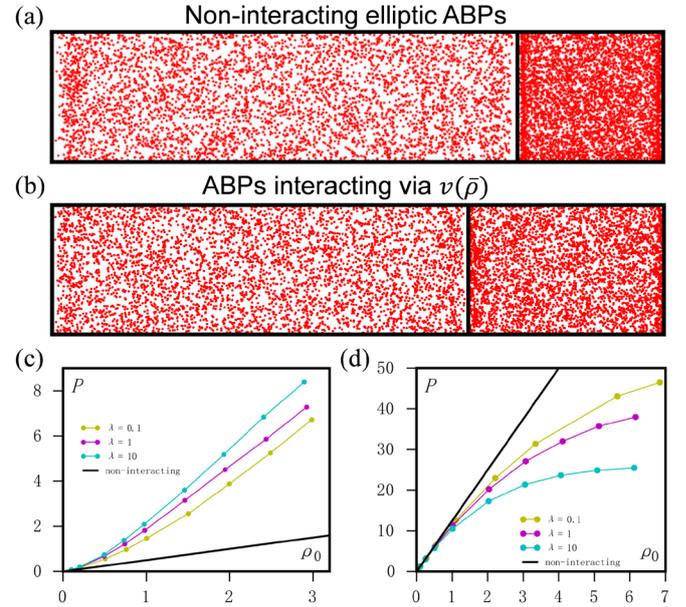

Figure 8. Test of the existence of an equation of state. Steady-state configurations of 10,000 ABPs in a $200 \times 50$ cavity divided by a mobile asymmetric harmonic wall ($\lambda = 1$ on the left, $\lambda = 4$ on the right, where $\lambda$ denotes the stiffness of the asymmetric harmonic wall). A spontaneous compression of the right half of the system indicates the absence of a well-defined equation of state. (a) Non-interacting elliptical ABPs. (b) ABPs with density-dependent propulsion speed. Pressure–density relations for interacting self-propelled particles. (c) Aligning ABPs. Pressure curves obtained for different wall stiffness values $\lambda$ separate clearly, indicating that the measured pressure depends on wall properties and that no well-defined equation of state exists for this system. (d) Quorum-sensing ABPs. The equation of state breaks down when the propulsion speed decreases with local density. Reproduced from Ref.[32], with permission from Springer Nature, ©2015.

For confined quorum sensing particles (Figure 8d), *i.e.*, a speed reduction according to the local swimmer density as is the case for some bacteria, destroys the equation of state[32] (Figure 8b). For active Brownian particles with quorum-sensing interactions, the propulsion speed $v(\bar{\rho})$ depends explicitly on the coarse-grained local density $\bar{\rho}(r)$. By applying Itô calculus to the density field $P(r,\theta)$, the mechanical pressure can be expressed in a microscopic form:

$$P = \zeta D_0 n_0 + \frac{\zeta}{2D_R}\langle v(\bar{\rho})^2(\hat{\rho}+m_2)\rangle_0 + \cdots \quad (37)$$

Where $\hat{\rho}$ and $m_2$ denote angular moments of $P$. This exact expression shows that $P$ is determined not only by bulk variables such as $n_0$ and the average propulsion speed, but also by spatial variations of $v(\bar{\rho})$ near boundaries. The coupling between motility and local density modifies the momentum flux at the walls, thereby breaking the equation of state in systems with density-dependent propulsion.

It is worth noting that, unlike the $U_\phi$ we discussed in section 3.1.1 which reflects the microscopic statistical configuration of the system but is not itself a measure of environmental sensing or interaction. In contrast, quorum sensing is an interaction mechanism widely observed in biological and active matter systems, in which particles detect and respond to local density via chemical signaling. This process modifies dynamical parameters such as propulsion speed $U(\bar{\rho})$ or rotational diffusivity $D_R(\rho)$, thereby indirectly affecting swim pressure, but it is not equivalent to $U_\phi$.

### 4.3 Boundary-Induced effects

#### 4.3.1 Stiffness of the boundary

Boundary dependence of the effective active pressure can be further emphasized[86] by demonstrating that the stiffness of the confining wall affects the measured active pressure. In their model, particles interacting with soft versus stiff walls produce distinct steady-state pressures, even when bulk properties remain unchanged (Figure 9a). Wall deformation changes the local wall normal, altering particle–boundary momentum exchange. For anisotropic particles this acts as an external torque, while for spherical particles it induces reorientation through geometry, leading in both cases to a wall-stiffness dependence of active pressure.

As shown in the figure, different wall stiffness values alter displacement of the wall accordingly (Figure 9b). This behavior confirms that the measured pressure is not determined solely by bulk variables and that boundary mechanics can play an intrinsic role in stress generation. Their results provide direct evidence that, in active systems, the coupling between particle dynamics and boundary properties constitutes a genuine mechanism for the breakdown of the equation of state.

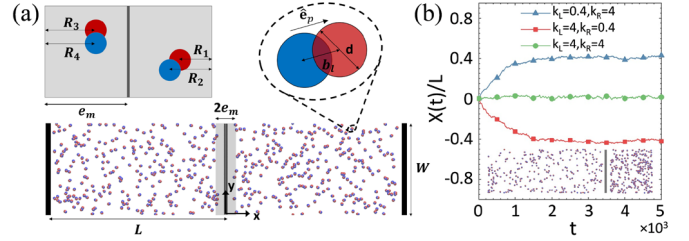

**Figure 9. Effect of wall stiffness on the motion of a mobile partition in active dumbbell systems.** (a) Simulation box confined by two fixed walls in the $x$ direction with a mobile wall separating dumbbells into two compartments. Reproduced from Ref.[86], *Scientific Reports*, 2021, under Creative Commons CC BY 4.0 license. (b) The time evolution of the mobile wall position was tracked under three wall stiffness combinations. Shown inset is the final configuration for $k_L = 0.4$ and $k_R = 4$. Results are averaged over five runs with different initial dumbbell distributions. Reproduced from Ref.[86], *Scientific Reports*, 2021, under Creative Commons CC BY 4.0 license.

#### 4.3.2 Alignment interaction between particle and wall

If torque is present when particles interact with boundaries, the pressure becomes boundary-dependent, and the EOS fails[38]. This is clearly demonstrated by simulations of active elliptical particles confined in two chambers[32], as shown in Figure 8a and 8b. Specifically, the effective pressure in a noninteracting Brownian active particle system confined by two walls along the $x$-direction exerting forces $-\nabla V(x)$ with a wall potential $V(x)$ can be expressed as

$$P = \frac{nU_0^2 \zeta \tau_R}{2} - \frac{U_0 \zeta_R \tau_R}{\zeta} \int dx \int d\theta\, \Gamma(x,\theta) \sin\theta\, \mathcal{P}(x,\theta), \quad (38)$$

where $\zeta_R$ is the rotational friction coefficient, $\mathcal{P}(x,\theta)$ is the probability distribution of finding a particle at distance from the wall $x$ and angle to the wall $\theta$.

$\Gamma(x,\theta)$ is an external force, *e.g.*, in order to model wall alignment of bacteria[32]. It becomes obvious now, that if $\Gamma(x,\theta) = 0$, the active swim pressure given earlier is recovered. However, if $\Gamma(x,\theta) \neq 0$, the pressure explicitly depends on the wall potential and thus an equation of state is violated. More intuitively, for non-interacting elliptical ABPs (torque present) (Figure 8a) and aligning ABPs (Figure 8c), the concept of an EOS fails.

Further, Marc Joyeux *et al.*[85] extends this perspective by analyzing underdamped self-propelled dumbbells confined in a two-dimensional chamber separated by a mobile wall. This model incorporates finite inertia and breaks the torque-free condition due to the asymmetric geometry of the dumbbells. The dumbbells exert asymmetric pressure on the mobile wall, causing it to shift from the center. In the overdamped limit, this displacement vanishes and the pressure distribution becomes symmetric. However, in the low-damping regime, the pressure response becomes more intricate and even exhibits non-monotonic behavior with respect to the damping coefficient. At low density and weak damping, resonant recollisions of high-angular-momentum particles near the wall lead to oscillatory pressure variations, which cannot be captured by an effective equation of state.

*4.3.3 Boundary geometry*

In the strong confinement limit, where the container size is smaller than the particle persistence length, active particles are almost entirely localized at the boundary, and their density and pressure profiles are directly dictated by the boundary geometry[88, 89]. For smooth convex boundaries, the steady-state density is proportional to the local curvature, leading to pronounced accumulation in high-curvature regions. Based on these findings, many simulations avoid such corner-trapping effects by choosing confinement geometries with rounded corners when setting up the simulation box[85, 90].

In polygonal geometries, where curvature is zero along edges and infinite at corners, particles become trapped at corners, producing sharp density peaks whose magnitude scales with the corner opening angle, while edges host almost no particles[37]. Although the theory is quantitatively most accurate for smoothly curved boundaries, it qualitatively captures the corner-trapping effect and the overall curvature dependence over a broad range of parameters. These results highlight that tailoring the curvature distribution of the confining boundary provides a direct route to control density and pressure patterns in active systems[88].

Taken together, these studies outline a coherent understanding of the conditions under which the equation of state fails in active systems. Factors such as torque arising from particle reorientation, interactions with structured boundaries, or external fields can disrupt the state function character of pressure. Similarly, density-dependent motility and collision-induced slow-down modify momentum transport in ways that are absent in equilibrium fluids, further undermining the applicability of a universal equation of state. The specific form of this breakdown is determined by the origin of torque, the density dependence of propulsion, the nature of particle dynamics (run-and-tumble or Brownian), and the imposed boundary conditions. Recognizing the interplay among these mechanisms is essential for capturing the diversity of active pressure phenomena and for assessing the prospects of a general thermodynamic framework for active matter.

*4.3.4 Experimental evidence of boundary effect*

To measure the pressure in active matter systems, Junot and his colleagues[91] placed self-propelled and passive isotropic disks on a two-dimensional vibrated plate, with a deformable chain consisting of jointed beads (acting as a membrane) separating the two sides. As the disks move, they exert forces on the chain, causing measurable deformations, which serve as a proxy for mechanical pressure (see Figure 10).

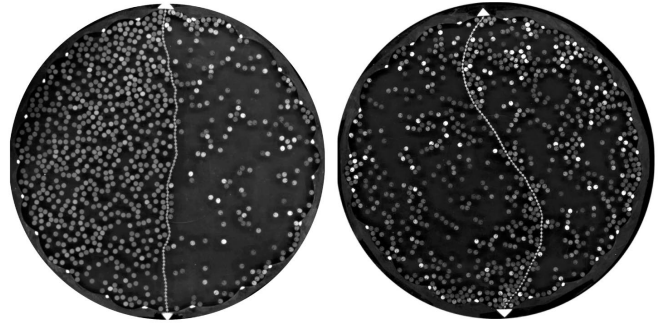

**Figure 10. Mechanical response of an active–passive mixture across a flexible membrane.** Left: Stable configuration with more passive particles on the left balancing fewer active disks on the right. Right: Membrane deformation under equal partitioning of active disks, indicating mechanical instability. Reproduced from Ref. [91], with permission from the American Physical Society.

In a configuration where passive and active disks were placed on opposite sides of the chain, by adjusting the number of disks on each side, the stresses exerted on the two sides of the membrane can be balanced, indicating that the effective pressure on both sides is equal and resulting in the chain aligning into a straight line. However, replacing the chain with a different dynamical response, such as changing the bead-size, the given balance is broken and the densities on both sides need to be readjusted, further demonstrating the sensitivity of active pressure to the probe's microscopic features. When both sides contained self-propelled disks, mechanical equilibrium could only be achieved under specific conditions — namely, when the number of disks on both sides was equal — resulting in an S-shaped chain. In all other configurations, the measured pressure was found to be dependent on the chain's dynamic properties, such as its relaxation time and responsiveness to particle collisions.

Although the chain-particle interactions in Ref.[91] are torque-free by construction, the S-shaped membrane instabaility does not necessarily imply torque-induced pressure differences. Instead, it may arise from curvature-dependent particle accumulation, similar to the wall curvature effects discussed in Ref.[88]. Meanwhile, for vibration-driven polar disks, the propulsive force is coupled with their orientational dynamics, also leads to the failure of EOS in this experiment.

Importantly, the experiment does not directly measure the bulk pressure nor verify its equivalence with the mechanical wall pressure. This result does not harm the concept of an equation of state for active matter with pressure as a state variable, but shows that this effective pressure does not necessarily equal the average wall force. This represents a significant research gap, calling for new experimental designs that can simultaneously access both bulk and boundary pressure in active matter systems.

## 5. Summary and perspective

This review has presented a comprehensive summary of recent theoretical and experimental progress in understanding active pressure, with particular emphasis on Active Brownian Particles. From the definition of swim stress in dilute suspensions to the breakdown of the equation of state in systems dominated by torques or inertial effects, we have discussed the current understanding of how nonequilibrium momentum flux translates into mechanical pressure. While significant advances have been made in elucidating microscopic mechanisms such as confinement effects, interparticle interactions, and boundary influences, general principles governing active pressure remain elusive.

Several key questions are yet to be fully resolved. For instance, to what extent can pressure in active systems be treated as a state function? Is it possible to develop a unified theoretical framework that reconciles boundary-sensitive pressures observed experimentally with bulk momentum flux–based descriptions from simulations? Recent studies suggest that the local pressure in active systems should exclude the swim pressure, with the self-propulsion force regarded as an external force generated by the surrounding environment[55, 83, 92, 93]. As partially discussed in section 3.2.1, while Speck *et al.*[55] showed that the swim stress was thought to vanish in the bulk, Das *et al.*[35] thought for more general active systems especially those with inhomogeneous activity distributions, and Steffenoni *et al.*[83] derived stress and pressure relations by considering the local momentum balance, proving the general lack of equivalance between the (local) bulk pressure and the pressure against confining walls, such that a universal definition of intrinsic pressure is not universally valid. Sun *et al.*[94] recent work establishes a more general intrinsic pressure framework and showing that it can recover mechanical equilibrium even in complex dry active systems with alignment, quorum sensing, and communication, marking a significant step toward a unified and versatile mechanical description of active matter.

Additionally, the roles of torque interactions and boundary geometry in disrupting or preserving an equation of state continue to be actively investigated, particularly in systems with inhomogeneous particle accumulation or phase coexistence[5, 32, 38, 71, 72, 79, 85, 86]. Experimental challenges further complicate this issue, as the measurement of swim pressure is often obscured by thermal noise[28, 54] or influenced by the specific properties of measurement techniques, hindering direct comparisons between theory and experiments.

A yet different kind of effective pressure resulting from activity is observed in chiral active systems, *i.e.*, systems consisting of active units showing not translational, but rotational acitve dynamics in a common direction[95]. Therein, an effective pressure arises proportional to a transport coefficient peculiar to chiral active systems called odd viscosity and pointing into the direction of emergent vorticity gradients leading to correlations between vorticity and density[96]. To what extent this effective pressure bears similarities to the active pressure concept presented in this review is still an open question.

Despite these complexities, active pressure has demonstrated considerable potential for practical applications. Recent studies have shown that active pressure is more than an abstract theoretical concept, it is a functional and controllable mechanical resource. Examples range from enabling directed cargo transport via topological edge currents[97] to harnessing nonequilibrium stress fluctuations for energy conversion in microscale engines[98]. Looking forward, establishing clearer connections among local stress distributions, particle-level interactions, and collective emergent behaviors will be critical for fully realizing the technological potential of active pressure in areas such as soft robotics, synthetic materials design, and bio-inspired micromachines.

## Acknowledgements

Y.G. acknowledges financial support from the General Program of the National Natural Science Foundation of China (12474195), the Key Project of Guangdong Provincial Department of Education (2023ZDZX3021), and Natural Science Foundation of Guangdong Province (2024A1515011343).